\documentclass[twocolumn]{aastex701}
\usepackage{savesym}
\savesymbol{tablenum}
\usepackage{siunitx}
\restoresymbol{SIX}{tablenum}
\usepackage{multirow}
\usepackage{threeparttable}
\usepackage{amsmath}
\usepackage{physics2}
\usephysicsmodule{ab.legacy}
\usephysicsmodule{ab}
\usephysicsmodule{braket}
\usepackage{bm}
\usepackage{derivative}

\shorttitle{Modeling Amorphous Dust Emission Based on the SP Model}
\shortauthors{Nashimoto}

\begin{document}

\title{Modeling Long-Wavelength Amorphous Dust Emission Based on the Physically Motivated Soft-Potential Model}

\correspondingauthor{Masashi Nashimoto}
\email{M.Nashimoto@niihama.kosen-ac.jp, nashimoto.astr@gmail.com}

\author[0000-0002-1221-1708]{Masashi Nashimoto}
\affiliation{National Institute of Technology, Niihama College, 7-1, Yagumo-cho, Niihama, Ehime 792-8580, Japan}
\email{M.Nashimoto@niihama.kosen-ac.jp, nashimoto.astr@gmail.com}

\begin{abstract}
We propose a new amorphous dust emission model based on the soft-potential (SP) model, applicable in the long-wavelength range from the far-infrared to the microwave. The SP model is widely accepted in material physics to explain amorphous thermal properties and is an extension of the two-level systems (TLS) model, which has been applied in interstellar amorphous dust physics. In the SP model, by assuming that some atoms composing amorphous dust are trapped in a double-well potential (DWP) described by a quartic function, the electric interaction can be solved directly, allowing the absorption cross-section of the amorphous dust to be calculated. We present numerical and analytical solutions for the absorption cross-section of amorphous dust and compare these results, finding good agreement for the DWP with a sufficiently high potential barrier. Our findings show that the SP model can reproduce the observed spectrum of the Perseus molecular cloud with slightly better accuracy than the conventional TLS model. Additionally, the SP model can more effectively explain various long-wavelength dust emission features, such as the spectrum flattening in the submillimeter range and anomalous microwave emission (AME), compared to the TLS model. Further comparison and verification with observational data and laboratory measurements are necessary to refine the model.
\end{abstract}

\keywords{Interstellar dust (836) --- Dust physics (2229) --- Dust continuum emission (412) --- Far infrared astronomy (529) --- Submillimeter astronomy (1647) --- Millimeter astronomy (1061)}

\section{Introduction}\label{sec:intro}

Large dust grains, found throughout interstellar space and primarily responsible for the majority of dust mass, are in thermal equilibrium with the interstellar radiation field (ISRF) and emit thermal photons at wavelengths longer than the far-infrared (FIR) range (e.g., \citealp{Draine03}). The primary components of large dust particles are mainly silicate and carbonaceous materials (e.g., \citealp{Draine&Lee84}). Comparisons between near-infrared (NIR) astronomical observations and laboratory measurements of silicate materials suggest that most interstellar silicate dust is amorphous rather than crystalline \citep{Kemper+04}. Carbonaceous materials with amorphous properties, such as amorphous carbons \citep{Duley+93} and hydrogenated amorphous carbons \citep{Jones+17}, are considered candidates for large carbonaceous dust grains. Amorphous properties are a universal characteristic of interstellar dust.

The semi-classical Lorentz and Drude models, which describe the optical properties of materials (c.f., \citealp{Bohren&Huffman83}), can explain laboratory measurements of the crystalline materials' optics. However, amorphous materials have been reported to exhibit behavior that these models cannot explain: the spectral index of the absorption coefficient, $\beta$, deviates from the expected value $\beta=2$ in both models and also varies with temperature $T$ (e.g., \citealp{Coupeaud+11, Demyk1+17, Demyk2+17}). It has been noted that there is an anti-correlation between $\beta$ and $T$ in amorphous materials \citep{Agladze+94, Agladze+96}, and astronomical observations have similarly reported such a trend in interstellar dust (e.g., \citealp{Dupac+03, Desert+08, Anderson+10, Paradis+10, Planck+11}). These findings suggest that the physics of amorphous materials must be considered in order to fully understand long-wavelength interstellar dust emission.

\cite{Meny+07} modeled the thermal emission from amorphous dust at wavelengths longer than the FIR using the two-level systems (TLS) model.  The TLS model, independently proposed by \cite{Anderson+72} and \cite{Phillips72}, explains the universal anomalies in the low-temperature thermal properties of amorphous materials (c.f., \citealp{Phillips87}). The amorphous dust emission model based on the TLS model has been shown to reproduce observations from FIR to millimeter range \citep{Paradis+11, Odegard+16}. \cite{Jones09} proposed that resonant transitions between the lowest two levels in the TLS model could explain anomalous microwave emission (AME), an unidentified interstellar radiation component peaking around \SI{30}{GHz} \citep{Kogut+96,Leitch+97,Davies+06}. \cite{Nashimoto+20pasj} enhanced the amorphous dust model proposed by \cite{Meny+07} and developed a more comprehensive model for amorphous dust emission from FIR to microwave wavelengths based on the TLS model. \cite{Meny+07} implicitly assumed a delta-function-type electric susceptibility for the resonance transitions of individual atoms in the amorphous dust, but \cite{Nashimoto+20pasj} applied the Lorentzian function, as derived from the governing equation, to optimize the TLS model for describing AME. Besides, in \cite{Nashimoto+20pasj}, the parameters of the TLS model were assumed to be distributed within a finite range rather than an infinite range, calculating in the same manner for each emission process. These modifications, ignored in \cite{Meny+07} in order to simplify the equation, gave significant changes in the spectrum when calculated more accurately. \cite{Nashimoto+20apjl} demonstrated that the TLS model could simultaneously describe the intensity and polarized spectra from FIR to microwave, including the AME, by considering a two-component amorphous dust model. The TLS model is expected to provide a unified description of dust emission across the FIR to microwave range.

The TLS model assumes the energy splitting caused by atoms in amorphous dust being trapped in a double-well potential (DWP), but it does not explicitly solve the Schr\"{o}dinger equation for a specific DWP. As a result, the parameters characterizing the TLS model, such as the energy difference between two levels, $\epsilon$, and the potential barrier height separating the DWP, $V_b$, are treated as independent variables. As $V_b$ increases, the states become more localized at the respective minima of the DWP, making energy splitting less likely and causing $\epsilon$ to decrease (e.g., \citealp{Landau&Lifshitz}). This implies that $\epsilon$ and $V_b$ are not entirely independent. In the TLS model, AME is explained as resonant transition emission between two levels, while the spectral flattening in the submillimeter (sub-mm) range is attributed to photons emitted when a disturbed state, affected by an incident electromagnetic wave, relaxes to thermal equilibrium \citep{Nashimoto+20pasj, Nashimoto+20apjl}. Resonant transitions depend solely on $\epsilon$ and are independent of $V_b$. In contrast, state relaxation caused by thermal hopping over the potential barrier explicitly depends on both $\epsilon$ and $V_b$. The relationship between $\epsilon$ and $V_b$ must be addressed precisely to verify whether both AME and sub-mm radiation can be consistently explained by amorphous dust. However, this critical relationship is not considered in the TLS model, leading to an inability to calculate amorphous dust radiation self-consistently from FIR to microwave wavelengths.

Although the TLS model can account for the amorphous properties of materials below approximately \SI{1}{K}, it fails to reproduce the behavior observed around \SI{10}{K}: such as the fifth power temperature dependence of the heat capacity and the plateau in the thermal conductivity \citep{Zeller&Pohl71}. For a detailed and comprehensive review on the amorphous properties, see \cite{Ramos+22} and references therein. To explain these measurements, the soft-potential (SP) model, an extension of the TLS model, was proposed \citep{Karpov+83, Ramos+93}. By describing the potential as a quartic function, the SP model incorporates both the TLS derived from the DWP and the anharmonicity introduced by the quartic term. With its ability to explain thermal properties around \SI{10}{K}, the SP model is expected to provide a universal explanation for the thermal anomalies of amorphous materials across a wide temperature range. However, the SP model has not yet been applied to interstellar dust physics.

In this paper, we apply the SP model to the optical properties of amorphous dust and revise the amorphous dust emission model. By assuming that the potential can be described by a quartic function, we solve the emission mechanisms arising from resonant transitions between two levels and state relaxation in a self-consistent manner. We also evaluate whether the SP model can explain the unique properties of dust emission from the FIR to microwave range, as effectively as the TLS model, within a unified framework. The structure of this paper is as follows: Sec.\,\ref{sec:SPmodel} details the SP model. In Sec.\,\ref{sec:Cabs}, we derive the absorption cross-sections of amorphous dust based on the SP model and demonstrate their frequency dependence. Sec.\,\ref{sec:CompObs} compares our amorphous dust model with observational data. In Sec.\,\ref{sec:discussion}, we discuss the improvements and predictions of our model. Finally, Sec.\,\ref{sec:conclusion} presents our conclusions.
\section{Soft-Potential Model} 
\label{sec:SPmodel}

In this section, we modeled the interaction between the atomic TLS and the electric field based on the SP model. When a quartic function describes the potential, it includes a single-well type, but we only considered a DWP here. Note that this assumption differs from the original SP model, which also takes into account the anharmonic single-well potential, and belongs to the TLS model category. To ensure clarity, however, we consistently name the model that considers the TLS derived from quartic DWPs as the SP model and the conventional model proposed by \cite{Anderson+72} and \cite{Phillips72} as the TLS model. We refer to the solution method in the standard TLS model, which assumes that states localized at each minimum of the DWP undergo energy splitting due to the tunneling effect, as the standard tunneling approximation (STA).

\subsection{Setup} 
\label{sec:setup}
Before constructing the model, we summarize assumptions for the properties of amorphous dust in this study. The TLS and SP models are expected to apply regardless of the amorphous dust species, such as silicate or carbon, since both are physical models introduced to describe the universality of amorphous materials. Differences in amorphous dust species are expected to be reflected only in the values of the model parameters. Given that amorphous materials are essentially randomly deformed crystalline structures, the assumption of a uniform isotropy in the internal structure of amorphous dust is reasonable. Since the SP model describes microscopic phenomena for atoms and macroscopic properties such as the size and shape of dust particles are not the essence of the SP model, for simplicity, we assume that the amorphous dust is spherical and its radius is \SI{0.1}{\micron}. This size is typical of interstellar dust particles, called large grains, and is small enough that the interaction with sub-mm waves can be accurately approximated by the Rayleigh scattering theory (the scattering contribution is sufficiently small compared to absorption). Besides, the effect of the dust surface is also negligible. For the electromagnetic response to amorphous dust, the influence of the magnetic field is negligibly small, and the influence of the electric field is approximated linearly. The nonlinear electric field response is not considered. The electric field is treated classically without quantization. By assuming the Lorentz local electric field, we assume that the local electric field is uniform within the dust particle and that each atom in the dust is in a similar local electric field (c.f., \citealp{Kittel04}). For simplicity, in the case of dust composed of two or more different chemical elements, assuming that a single pseudo-atom with the average physical properties of the constituent atoms forms amorphous dust, and we describe the physical quantity associated with the atoms constituting the dust as a single value for each dust species.

\subsection{Standard two-level systems model} 
\label{sec:TLS}
The SP model is an extension of the TLS model, which has much in common. In this section, we briefly summarize the TLS model and then clarify how the SP model improves the TLS model in the following sections. Refer to previous studies for more details on the TLS model (e.g., \citealp{Phillips87,Meny+07,Nashimoto+20pasj}).

When an atom is trapped in the DWP and forms the TLS, its ground and first excited states, $\ket{\varphi_1}$ and $\ket{\varphi_2}$, satisfy the time-independent Schr\"{o}dinger equation,
\begin{align}
    \hat{H}_0 \ket{\varphi_n} =
    \epsilon_n \ket{\varphi_n},
    \label{eq:SchrodingerEq-stable}
\end{align}
where $\hat{H}_0$ is the unperturbed Hamiltonian, and $\epsilon_n$ is the eigenenergy for the $n$-th order state (the ground and first excited states correspond to $n=1$ and 2, respectively). Assuming the STA, the localized states in each potential minima of the DWP, where $\ket{\phi_L}$ and $\ket{\phi_R}$ correspond to two minima $x_L$ and $x_R$ of the DWP ($x_L < x_R$), are expressed as,
\begin{alignat}{2}
    \ket{\varphi_1} &=
    \cos \theta \ket{\phi_L} &&- \sin \theta \ket{\phi_R},
    \label{eq:phi1} \\
    \ket{\varphi_2} &=
    \sin \theta \ket{\phi_L} &&+ \cos \theta \ket{\phi_R},
    \label{eq:phi2} \\
    \tan2\theta &\equiv \frac{\Delta_0}{\Delta},
\end{alignat}
where $\Delta$ and $\Delta_0$ represent the potential asymmetry and the tunneling energy, respectively, and are defined as,
\begin{align}
    \Delta &\equiv 
    \braket[3]{\phi_R}{\hat{H}_0}{\phi_R} - \braket[3]{\phi_L}{\hat{H}_0}{\phi_L},
    \label{eq:Delta} \\
    \Delta_0 &\equiv 
    2\vab{\braket[3]{\phi_L}{\hat{H}_0}{\phi_R}}.
    \label{eq:Delta0}
\end{align}
The energy difference between two levels, $\epsilon \equiv \epsilon_2 - \epsilon_1$, is given as,
\begin{align}
    \epsilon  =
    \sqrt{\Delta^2 + \Delta_0^2}.
    \label{eq:epsilon}
\end{align}

Supposing a system of $N_\mathrm{DWP}$ atoms (note that $N_\mathrm{DWP}$ refers to the number of atoms trapped in the DWP not the total number of atoms in the amorphous dust) and assuming that the ground and first excited states, $\ket{\varphi_1}$ and $\ket{\varphi_2}$, form a complete orthonormal system, any state for an $i$-th atom can be written in terms of complex coefficients $a_{n,i}$ as,
\begin{align}
    \ket{\psi_i(t)} = 
    a_{1,i}(t)\ket{\varphi_1} + a_{2,i}(t)\ket{\varphi_2}.
    \label{eq:psi}
\end{align}
The density operator $\hat{\rho}$ is defined as,
\begin{align}
    \hat{\rho} \equiv
    \frac{1}{N_\mathrm{DWP}} \sum_{i=1}^{N_\mathrm{DWP}} 
    \ketbra{\psi_i}{\psi_i}.
\end{align}
The matrix elements of the density operator are as,
\begin{align}
    \begin{pmatrix}
        \rho_{11} & \rho_{12} \\ 
        \rho_{21} & \rho_{22}
    \end{pmatrix} =
     \frac{1}{N_\mathrm{DWP}}
    \begin{pmatrix}
        \displaystyle{\sum_{i=1}^{N_\mathrm{DWP}}} a_{1,i}a_{1,i}^* & 
        \displaystyle{\sum_{i=1}^{N_\mathrm{DWP}}} a_{1,i}a_{2,i}^* \\
        \displaystyle{\sum_{i=1}^{N_\mathrm{DWP}}} a_{2,i}a_{1,i}^* & 
        \displaystyle{\sum_{i=1}^{N_\mathrm{DWP}}} a_{2,i}a_{2,i}^*
    \end{pmatrix},
\end{align}
where the symbol $*$ denotes complex conjugation. The Liouville-von Neumann equation describes the time evolution of the density matrix as,
\begin{align}
    i\hbar \pdv*{\hat{\rho}}{t} &=
    \bab{\hat{H}, \hat{\rho}},
    \label{eq:Liouville}
\end{align}
where square brackets are the commutator. The Hamiltonian is given as $\hat{H} = \hat{H}_0 + \hat{H}_\mathrm{ele}$, where $\hat{H}_\mathrm{ele}$ is the electric interaction Hamiltonian defined as,
\begin{align}
    \hat{H}_\mathrm{ele}(t) = 
    -q \bm{E}_\mathrm{loc}(t) \cdot \hat{\bm{r}},  
    \label{eq:Hamiltonian_ele}
\end{align}
where $\bm{E}_\mathrm{loc}$ is the local electric field, and $q$ and $\hat{\bm{r}}$ are the atomic charge and position operator, respectively. We assume that the interaction energy due to the local electric field is sufficiently small compared to the eigenenergies and that the contribution from $\hat{H}_\mathrm{ele}$ is perturbatively solvable.

Introducing the Bloch vector defined as,
\begin{align}
    \begin{pmatrix}
        u \\ v \\ w
    \end{pmatrix} &\equiv 
    \begin{pmatrix}
        \rho_{21} + \rho_{12} \\
        i\pab{\rho_{21} - \rho_{12}} \\
        \rho_{22} - \rho_{11}
    \end{pmatrix},
    \label{eq:BlochVec} \\
    u_\pm &\equiv u \pm i v .
\end{align}
By solving the Bloch equation obtained by substituting the Bloch vector into the Liouville-von Neumann equation in Eq.\,\eqref{eq:Liouville}, each component of the Bloch vector is obtained as,
\begin{align}
    u_\pm^{(\omega)} &=
    \frac{\pm2}{i\hbar} 
    \frac{q\bm{E}_\mathrm{loc}^{(\omega)}\cdot \bm{r}_{12}}
    {\gamma-i\pab{\omega\pm\omega_0}}
    \tanh\pab{\frac{\epsilon}{2k_\mathrm{B}T}},
    \label{eq:u_pm} \\ 
    w^{(\omega)} &=
    \frac{\Gamma}{\Gamma-i\omega} 
    \frac{q \bm{E}_\mathrm{loc}^{(\omega)} \cdot \pab{\bm{r}_{22} - \bm{r}_{11}}}
    {2k_\mathrm{B}T} 
    \mathrm{sech}^2 \pab{\frac{\epsilon}{2k_\mathrm{B}T}},
    \label{eq:w} \\ 
    \bm{r}_{nm} &\equiv
    \braket[3]{\varphi_n}{\hat{\bm{r}}}{\varphi_m},
    \label{eq:r_nm}
\end{align}
where $\bm{E}_\mathrm{loc}(t) = \bm{E}_\mathrm{loc}^{(\omega)}e^{-i\omega t} + \bm{E}_\mathrm{loc}^{(-\omega)}e^{i\omega t}$, $u_\pm(t) = u_\pm^{(\omega)}e^{\mp i\omega t}$, $w(t) = w^{(\omega)}e^{-i\omega t} + w^{(-\omega)}e^{i\omega t}$, and $\gamma$ and $\Gamma$ represent the phase and state relaxation rates of the Bloch vector, respectively. These relaxation rates originate from the phenomenological addition of relaxation terms to the Bloch equation, meaning relaxing to a thermal equilibrium state at temperature $T$. The state relaxation is contributed by the quantum effect originating from the tunneling process driven by the interaction between the atomic TLS and the phonon field (tunneling relaxation) and the classical one caused by hopping over the potential barrier (hopping relaxation). The respective state relaxation rates, $\Gamma_\mathrm{tun}$ and $\Gamma_\mathrm{hop}$, are given as,
\begin{align}
    \Gamma_\mathrm{tun} &=
    \pab{\frac{2\Lambda_t^2}{c_t^5} + \frac{\Lambda_l^2}{c_l^5}}
    \frac{\epsilon^3}{2\pi\hbar^4\varrho}
    \vab{\frac{\bm{r}_{12}}{x_0}}^2
    \coth \pab{\frac{\epsilon}{2k_\mathrm{B}T}},
    \label{eq:Gamma-tun} \\
    \Gamma_\mathrm{hop} &= 
    \Gamma_\mathrm{hop}^0 \exp\pab{-\frac{V_b}{k_\mathrm{B}T}},
    \label{eq:Gamma-hop}
\end{align}
where $c_{t(l)}$ are the sound velocities of transverse (longitudinal) elastic waves, $\Lambda_{t(l)}$ are the coupling constants of the elastic dipole for transverse (longitudinal) waves, $\varrho$ is the dust mass density, $x_0$ is the normalized constant for atomic displacement, and $\Gamma_\mathrm{hop}^0$ is the attempt frequency, which have different values for each material (see Tab.\,\ref{tab:parameters}).

The matrix components of the position operator are described as,
\begin{align}
    \begin{pmatrix}
        \bm{r}_{11} & \bm{r}_{12} \\
        \bm{r}_{21} & \bm{r}_{22}
    \end{pmatrix} &=
    \begin{pmatrix}
        -\bm{r}_0\cos2\theta & -\bm{r}_0\sin2\theta \\
        -\bm{r}_0\sin2\theta &  \bm{r}_0\cos2\theta
    \end{pmatrix},
    \label{eq:r_nm_approx} \\
    \bm{r}_0 &\equiv
    \frac{\braket[3]{\phi_R}{\hat{\bm{r}}}{\phi_R}
    -\braket[3]{\phi_L}{\hat{\bm{r}}}{\phi_L}}{2},
    \label{eq:r0}
\end{align}
where the crossing terms are ignored ($\braket[3]{\phi_L}{\hat{\bm{r}}}{\phi_R} = \braket[3]{\phi_R}{\hat{\bm{r}}}{\phi_L} = \bm{0}$). In practice, $\bm{r}_0$ depends on the potential shape, as seen in Eq.\,\eqref{eq:r0}, but the TLS model treats the absolute value of $\bm{r}_0$ as the parameter $x_0$. Therefore, the TLS model can be described by regarding $\Delta$, $\Delta_0$, and $V_b$ as independent variables, so there is no need to calculate the localized states, $\ket{\phi_L}$ and $\ket{\phi_R}$.

\subsection{Basis of the soft-potential model} 
\label{sec:basis-SPmodel}
We consider a situation in which some atoms constituting amorphous dust are trapped in a DWP described by a quartic function. Then, we solve the interaction between the atomic TLS caused by the DWP and the electric field.

\subsubsection{Quartic double-well potential} 
\begin{figure*}[t]
\centering
\includegraphics[scale=1]{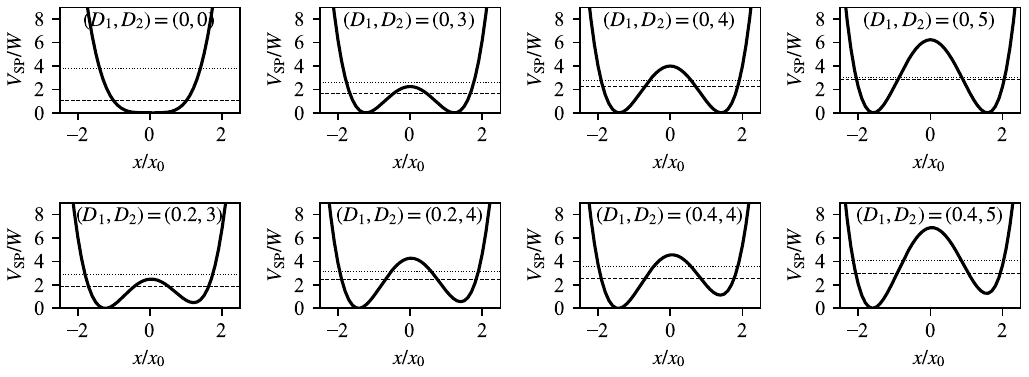}
\caption{
    Dependence of the parameters, $D_1$ and $D_2$, on the shape of the potential $V_\mathrm{SP}$. The potentials are shifted to regard the potential's minimum value as zero. The upper and lower panels show symmetric ($D_1=0$) and asymmetric ($D_1>0$) case, respectively. The dashed (dotted) lines in each panel indicate the ground (first excited) energy corresponding to each potential calculated numerically.
}
\label{fig:Vsp}
\end{figure*}

The Hamiltonian $\hat{H}_0$, for the eigenstate of an atom with the mass $m$ for the potential $V_\mathrm{SP}$, is given as,
\begin{align}
    \hat{H}_0 =
    \frac{\hat{p}^2}{2m} + V_\mathrm{SP}(x),
    \label{eq:Hamiltonian0}
\end{align}
where $\hat{p}$ is the momentum operator. The potential $V_\mathrm{SP}$ is defined as,
\begin{align}
    V_\mathrm{SP}(x) \equiv 
    W \pab{ 
        D_1\pab{\frac{x}{x_0}} 
        - D_2\pab{\frac{x}{x_0}}^2 
        + \pab{\frac{x}{x_0}}^4 } 
    , \label{eq:Vsp}
\end{align}
where the coordinate system is assumed in which the cubic term is zero. The dimensionless parameters, $D_1$ and $D_2$, represent the asymmetry of the potential and the magnitude of the restoring force, respectively. Fig.\,\ref{fig:Vsp} shows the dependence of the dimensionless parameters on the shape of the potential $V_\mathrm{SP}$. As $D_1$ increases, the asymmetry of the potential increases; as $D_2$ increases, the potential becomes a deeper well. In previous studies (e.g., \citealp{Karpov+83,Ramos+93}), the quadratic term in $V_\mathrm{SP}$ is defined with positive signs. Since we only treat the range where the quadratic terms are negative in this study (see Sec.\,\ref{sec:dist-func}), we define the coefficient $D_2$ with negative signs to treat $D_2$ as a positive parameter. The assumption that the internal structure of amorphous dust is isotropic implies that $D_1$ has even symmetry. The scale parameters $W$ and $x_0$ have dimensions of energy and distance and are about the atomic binding energy and lattice constant, respectively. When both parameters satisfy $W=\hbar^2/(2mx_0^2)$, one of $W$, $x_0$, and $m$ can be regarded as the dependent variable. The energy scale parameter $W$ also roughly corresponds to the conversion energy between TLS-derived tunneling modes and anharmonicity-derived vibrational modes in atomic systems. Fig.\,\ref{fig:Vsp} also shows numerical solutions of $\epsilon_1$ and $\epsilon_2$, and indicates that the energy difference $\epsilon$ basically increases as $D_1$ increases or $D_2$ decreases.

The potential barrier height $V_b$ is defined as,
\begin{align}
    V_b \equiv
    \frac{V_\mathrm{SP}(x_O) - V_\mathrm{SP}(x_L)}{2} 
    + \frac{V_\mathrm{SP}(x_O) - V_\mathrm{SP}(x_R)}{2},
    \label{eq:Vb}
\end{align}
where $x_O$ is the displacement of the maximum extremum point.

\subsubsection{Interaction with electric field}
The next step is to solve for the interaction between the atom and the electric field for the potential $V_\mathrm{SP}$. The solution of the Bloch vector can be written by Eqs.\,\eqref{eq:u_pm} and \eqref{eq:w} in the SP model as in the TLS model, but arguments for the SP model are $D_1$, $D_2$, and $W$ instead of $\Delta$, $\Delta_0$, and $V_b$. The TLS model describes the arbitrary state by assuming the STA and introducing localized states $\ket{\phi_L}$ and $\ket{\phi_R}$. In the SP model, on the other hand, the eigenstates and the eigenenegies can be obtained directly without assuming the STA by solving the Schr\"{o}dinger equation in Eq.\,\eqref{eq:SchrodingerEq-stable} using the Hamiltonian defined in Eq.\,\eqref{eq:Hamiltonian0}, and we can numerically calculate $\epsilon$ and $\bm{r}_{nm}$ without the approximations given by Eqs.\,\eqref{eq:epsilon} and \eqref{eq:r_nm_approx} according to the definitions.

\subsection{Analytical solution with the standard tunneling approximation} 
\label{sec:SP_ana}
The Schr\"{o}dinger equation and the Liouville-von Neumann equation for the potential described in Eq.\,\eqref{eq:Vsp} generally do not have elementary solutions, and no exact analytical solutions are available. According to \cite{Ramos+93}, we assume that the asymmetry of the potential is slight ($D_1 \ll 1$) and give an analytical approximate solution of the SP model applying the STA. In that case, we can apply the same formula in Sec.\,\ref{sec:TLS} as in the TLS model. For the variables $\Delta$, $\Delta_0$, and $V_b$, which are the arguments in the STA, the analytical approximate solutions are given as \citep{Ramos+93},
\begin{align}
    \Delta &=
    W D_1 \sqrt{2(D_2-1)},
    \label{eq:Delta_approx} \\
    \Delta_0 &=
    W D_2^{3/2} \exp \pab{1-\frac{\sqrt{2}}{3}D_2^{3/2}},
    \label{eq:Delta0_approx} \\
    V_b &=
    W \pab{\frac{D_2}{2}}^2.
    \label{eq:Vb_approx}
\end{align}
The approximation formula in Eq.\,\eqref{eq:r_nm_approx} can be used for the matrix component of the position operator, $\bm{r}_{nm}$, as in the TLS model. While the TLS model treats $\bm{r}_0$ as a parameter, the SP model regards as a variable described as,
\begin{align}
    \vab{\bm{r}_0} = 
    x_0 \sqrt{\frac{D_2-1}{2}}.
    \label{eq:r0_approx}
\end{align}
Eqs.\,\eqref{eq:Delta_approx} and \eqref{eq:r0_approx} give imaginary values when $D_2 < 1$, which is unsuitable. Besides, when $D_2 = 1$, the induced dipole moment (see Eq.\,\eqref{eq:dipole-ind}) is zero because $\vab{\bm{r}_0} = 0$, which is also excluded. Therefore, we consider the range $D_2 > 1$ when using the analytical solution.

\subsection{Comparison between numerical and analytical solutions}
\label{sec:CompNumAna}

\subsubsection{Energy difference}
\begin{figure}[t]
\centering
\includegraphics[scale=1]{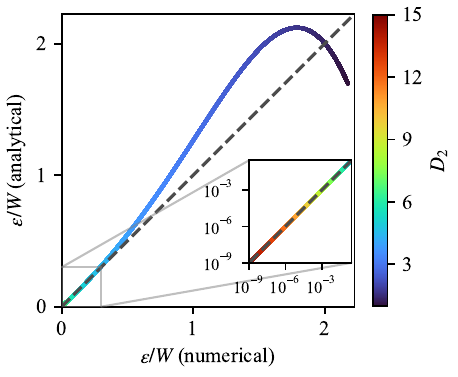}
\caption{
    Comparison of the computation of $\epsilon$ for $D_1=0$. The horizontal and vertical axes show the numerical and analytical solutions, respectively. The inset shows the enlargement on a logarithmic scale for $\epsilon/W \leq 0.3$. The dashed line indicates the case where both solutions coincide.
}
\label{fig:E_comp}
\end{figure}
\begin{figure*}[!t]
\centering
\includegraphics[scale=1]{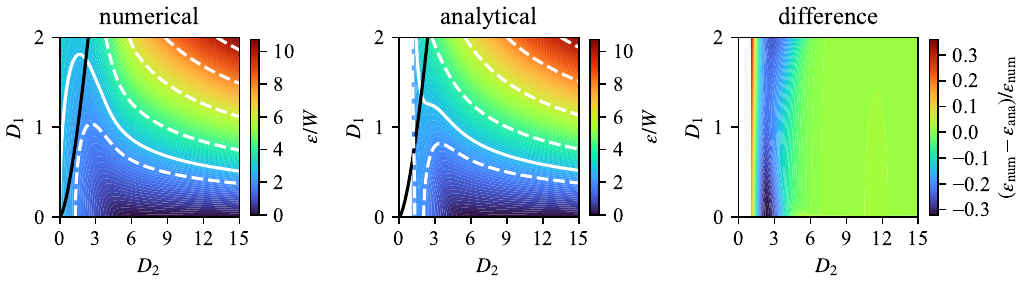}
\caption{
    Contour plots of energy difference in the $D_1$--$D_2$ plane. The left and middle figures show the results of numerical and analytical solutions, respectively. The solid and dashed white curves represent $\epsilon = \epsilon_\mathrm{max} \simeq 2.74W$ and $\epsilon/W =$ 2, 4, 6, 8, and 10, respectively. The potential is a double-well type in the region below the solid black curves (see Eq.\,\eqref{eq:condition-D2}). Since the analytical method does not yield real roots for $D_2 < 1$ (see Eq.\,\eqref{eq:Delta_approx}), that region is blank in the middle panel. The right figure shows the difference between the two solutions normalized by the numerical solution.
}
\label{fig:E_contour}
\end{figure*}

Fig.\,\ref{fig:E_comp} compares numerical and analytical solutions for $\epsilon$ when the potential is symmetric ($D_1=0$). Both solutions agree well for $\epsilon/W \lesssim 0.5$ (i.e., $D_2 \gtrsim 4$). On the other hand, the discrepancy gets larger as $D_2$ decreases. This is because the localized states at each minimum point of the DWP assumed in the analytical solution break down as the height of the potential barrier decreases with decreasing $D_2$.

Fig.\,\ref{fig:E_contour} shows the contour plots of the energy difference $\epsilon$ in the $D_1$--$D_2$ plane. As in the case of symmetric potential (i.e., $D_1=0$) demonstrated in Fig.\,\ref{fig:E_comp}, both numerical and analytical solutions agree well in the range $D_2 \gtrsim 4$. For $D_2 \lesssim 4$, in contrast, there is a discrepancy of up to 30\% between the two solutions. The distribution of the discrepancy depends mainly on $D_2$ and is relatively uniform for $D_1$.

\subsubsection{Potential barrier height}
\begin{figure}[t]
\centering
\includegraphics[scale=1]{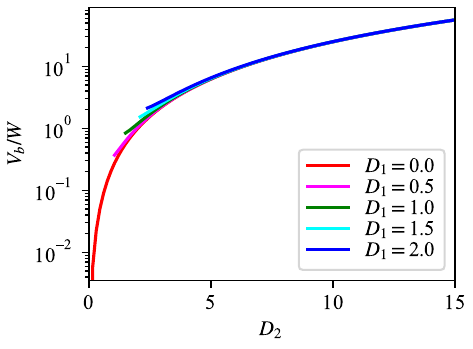}
\caption{
    Dependence of the potential barrier $V_b$ on $D_2$. 
    \added{
    This figure shows only the numerical solution since the analytical solution does not depend on $D_1$ and is identical to the numerical solution for $D_1=0$.} The curves are broken because $V_b$ cannot be defined in a part of $D_1$--$D_2$ plane (see text).
}
\label{fig:Vb-D2}
\end{figure}
Fig.\,\ref{fig:Vb-D2} shows the result of the numerical calculation of $V_b$ using Eq.\,\eqref{eq:Vb}. The analytical approximate solution is exact when the potential is symmetric, that is, $D_1=0$ (see Eq.\,\eqref{eq:Vb_approx}), and the numerical and analytical approximate solutions are identical. In the analytical approximate solution method, the dependence of $V_b$ on $D_1$ is ignored, so the solution when $D_1=0$ applies to all $D_1$. For $D_2 \gtrsim 4$, the numerical solution of $V_b$ and the analytical approximate solution coincide even for large values of $D_1$. For $D_2 \lesssim 4$, there is a difference between the two results, but the discrepancy is slight. The reason why $V_b$ is not calculated for small values of $D_2$ with $D_1 \neq 0$ is that the potential is not a double-well type, and therefore, the potential barrier cannot be defined (see Eq.\,\eqref{eq:condition-D2}).

\subsubsection{Matrix elements for displacement}
\begin{figure}[t]
\centering
\includegraphics[scale=1]{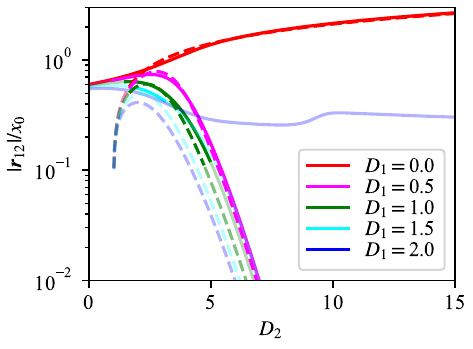}
\caption{
    Dependence of $\bm{r}_{12}$ on $D_2$. The absolute values are taken and normalized by $x_0$. The solid and dashed curves indicate numerical and analytical solutions, respectively. The lighter-colored sections of the curves mean the ranges not addressed in this paper (see Sec.\,\ref{sec:dist-func}).
}
\label{fig:r12}
\end{figure}
Fig.\,\ref{fig:r12} shows the results of computing $\bm{r}_{12}$ using Eqs.\,\eqref{eq:r_nm}, \eqref{eq:r_nm_approx} and \eqref{eq:r0_approx}. In the range of $D_1 \leq 0.5$, the results of both solution methods agree well, regardless of the value of $D_2$. For $D_1 \geq 1$, there are discrepancies, and the approximate shape of the curves is entirely different, especially for $D_1 = 2$. Since the SP model treated in this study targets only a limited region on the $D_1$--$D_2$ plane (see Sec.\,\ref{sec:dist-func}), it can be concluded that the analytical approximate solution is a good approximation to the numerical solution even within $D_1 \gtrsim 1$ when the comparison is limited to that range. 
\section{Modeling Amorphous Dust Absorption Cross-Sections}
\label{sec:Cabs}
The absorption cross-section of amorphous dust is derived from the interaction between atoms in amorphous dust and an electric field, which was derived in the previous section. The absorption cross-section of spherical dust, $C_\mathrm{abs}$, is given using the Rayleigh scattering theory as (e.g., \citealp{Bohren&Huffman83}),
\begin{align}
    C_\mathrm{abs} =
    \frac{3 \omega V}{c} \mathrm{Im} \pab{\frac{\varepsilon-1}{\varepsilon+2}},
    \label{eq:Cabs}
\end{align}
where $c$ is the speed of light, $V$ is the dust volume, $\varepsilon$ is the complex permittivity of the dust, and $\mathrm{Im}$ is a function that returns the imaginary part of the argument. In this section, the complex permittivity $\varepsilon$, a macroscopic physical quantity, is calculated based on the SP model, a microscopic physical model.

\subsection{Atomic polarizability}
The Clausius-Mossotti relationship, which relates the complex permittivity of amorphous dust, $\varepsilon$, to the atomic polarizability of the atoms consisting of the amorphous dust, $\alpha_\mathrm{SP}$, is expressed as (c.f., \citealp{Jackson62}),
\begin{align}
    \frac{\varepsilon-1}{\varepsilon+2} &=
    \frac{4\pi}{3} \frac{\alpha_\mathrm{SP}}{V},
    \label{eq:Clausius-Mossotti}
\end{align}
where $\alpha_\mathrm{SP}$ is the sum of the polarizabilities for each atom in the amorphous dust,
\begin{align}
    \alpha_\mathrm{SP} &\equiv
    \sum_{i=1}^{N_\mathrm{DWP}} \alpha_\mathrm{SP}^i.
\end{align}
Each atomic polarizability $\alpha_\mathrm{SP}^i$ is given by,
\begin{align}
    \bm{d}_\mathrm{ind}^i(\omega) = 
    \alpha_\mathrm{SP}^i \bm{E}_\mathrm{loc}^{(\omega)},
    \label{eq:polarizability}
\end{align}
where the local electric field is assumed to be uniform regardless of positions. The expected value for the induced dipole moment of $i$-th atom, $\bm{d}_\mathrm{ind}^i$, is also written as,
\begin{align}
    \bm{d}_\mathrm{ind}^i(\omega) &=
    q\pab{u_+^{(\omega)} + u_-^{(\omega)}} \bm{r}_{12}
    + \frac{q}{2}w^{(\omega)} \pab{\bm{r}_{22}-\bm{r}_{11}},
    \label{eq:dipole-ind} 
\end{align}
where the first and second terms are due to resonant transitions and state relaxation, respectively. As in previous studies \citep{Meny+07, Nashimoto+20pasj, Nashimoto+20apjl}, the state relaxation term is described by the sum of the tunneling and hopping terms (i.e., $w^{(\omega)} = w_\mathrm{tun}^{(\omega)} + w_\mathrm{hop}^{(\omega)}$).

Distribution functions of the model parameters are required to calculate the sum of the atomic polarizabilities. The standard assumption of the SP model in \cite{Ramos+93} is that $W$ has a single value in an amorphous material, and, in contrast, $D_1$ and $D_2$ are distributed across an arbitrary range based on the distribution function $f(D_1,D_2)$. Following this assumption, $\alpha_\mathrm{SP}$ is calculated by integrating $D_1$ and $D_2$ using $f(D_1,D_2)$ as,
\begin{align}
    \alpha_\mathrm{SP} =
    N_\mathrm{DWP} \iint \odif{D_1}\odif{D_2} f(D_1,D_2) 
    \alpha_\mathrm{SP}^i(D_1,D_2).
    \label{eq:alp_sp_int}
\end{align}
Based on the assumption of the isotropic internal structure of amorphous dust, the angle between $\bm{r}_{nm}$ and $\bm{E}_\mathrm{loc}^{(\omega)}$, $\vartheta$, is randomly distributed in the amorphous dust, so it is supposed to be $\cos^2\vartheta = 1/3$ by taking a spatial average. The following analysis uses a parameter, $f_\mathrm{SP}$, defined as,
\begin{align}
    f_\mathrm{SP} \equiv \frac{N_\mathrm{DWP}}{N_\mathrm{atom}},
    \label{eq:f_sp}
\end{align}
where $N_\mathrm{atom}$ is the total number of atoms in the dust and $N_\mathrm{DWP}$ means the number of atoms trapped in the DWP.

\subsection{Distribution function}
\label{sec:dist-func}
We assume a uniform distribution for $f(D_1,D_2)$ as in the previous study (e.g., \citealp{Ramos+93}),
\begin{align}
    f(D_1,D_2) = P_\mathrm{SP},
\end{align}
where $P_\mathrm{SP}$ is the dimensionless normalization constant of the distribution function, which is uniquely determined by the integral range of $D_1$ and $D_2$. In this paper, since we focus only on the contribution from the DWP, not considering the anharmonic single-well potential, we impose the necessary and sufficient condition that $V_\mathrm{SP}$ is a double-well type; in other words, $V_\mathrm{SP}$ has two extreme minima. In this case, the discriminant of the first derivative of $V_\mathrm{sp}$ must be positive. Therefore, we can restrict $D_1$ and $D_2$ as, 
\begin{align}
    D_1^2 < \pab{\frac{2}{3} D_2}^3.
    \label{eq:condition-D2}
\end{align}
When the asymmetry of the DWP is significant, the potential can be approximated as a single well for the first excited state and below, so almost no energy splitting occurs. To exclude such a singularly shaped potential, we consider only the range where $\epsilon$ is smaller than the energy difference at $D_1=D_2=0$, which is the maximum value and is numerically calculated as $\epsilon_{\max} = 2.74W$. The lower limits of $D_1$ and $D_2$ are set to zero, and the upper limit of $D_2$, $D_2^\mathrm{max}$, is a free parameter. The upper boundary for $D_1$ is determined by the inequalities in Eq.\,\eqref{eq:condition-D2} and $\epsilon \leq \epsilon_\mathrm{max}$ (see Fig.\,\ref{fig:E_contour}). As $D_2$ decreases, the energy splitting becomes smaller because the potential shape approaches a single-well type from a double-well type (see Fig.\,\ref{fig:Vsp}). Therefore, $\Delta_0$ is considered to decrease monotonically with respect to $D_2$. The range of $D_2$ satisfying this condition is as,
\begin{align}
    D_2 \geq \pab{\frac{9}{2}}^{1/3} \simeq 1.65.
    \label{eq:D2_min}
\end{align}
The inequality, $\epsilon \leq \epsilon_\mathrm{max}$, is equivalent to the following inequality in the analytical method using Eqs.\,\eqref{eq:Delta_approx} and \eqref{eq:Delta0_approx},
\begin{align}
    D_1 \leq
    \sqrt{\frac{\pab{\frac{\epsilon_\mathrm{max}}{W}}^2
    -D_2^3\exp\pab{2-\frac{1}{3}\pab{2D_2}^{3/2}}}{2\pab{D_2-1}}}.
    \label{eq:D1_max}
\end{align}

\begin{figure}[t]
    \centering
    \includegraphics[scale=1]{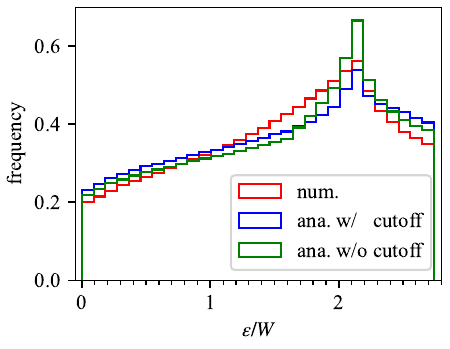}
    \caption{
    The frequency distributions of the energy difference for $D_2^\mathrm{max}=15$, normalized to have an integral value of 1. The red line indicates numerical solutions. The blue and green lines are analytical approximations with and without the cutoff in Eq.\,\eqref{eq:D2_min}, respectively.
    }
    \label{fig:E_frequency}
\end{figure}
Fig.\,\ref{fig:E_frequency} shows the frequency distribution of the energy difference $\epsilon$ calculated according to $f(D_1,D_2)$. The overall characteristics of the frequency distribution are the same, with a peak at $\epsilon/W \simeq 2.1$ for both solutions. The peak position of the distribution corresponds to $\epsilon \simeq \Delta_0^\mathrm{max}$, that is, the maximum energy difference in the range where the potential is symmetric. The fact that the analytical approximate solution is more to the right than the numerical solution is thought to reflect a discrepancy in the calculation results around $D_2 \simeq 3$, as seen in Fig.\,\ref{fig:E_contour}. When the analytical approximation is compared with and without the condition in Eq.\,\eqref{eq:D2_min}, the overall distribution is more concentrated around the peak when this condition is not taken into account. It can be concluded that the condition in Eq.\,\eqref{eq:D2_min} effectively asymptotes the analytical approximation closer to the numerical solution.

In short, $\alpha_\mathrm{SP}$ is calculated from Eq.\,\eqref{eq:alp_sp_int}, $\alpha_\mathrm{SP}^i$ from Eqs.\,\eqref{eq:polarizability} and \eqref{eq:dipole-ind}, and variables in Eq.\,\eqref{eq:dipole-ind} come from Eqs.\,\eqref{eq:u_pm}, \eqref{eq:w}, \eqref{eq:r_nm_approx}, \eqref{eq:r0}, and \eqref{eq:r0_approx} depending on the numerical or analytical case.

\subsection{Absorption coefficient} 
\label{sec:Qabs}
To demonstrate the SP model, we calculated the frequency dependence of the absorption coefficients, $Q_\mathrm{abs} \equiv C_\mathrm{abs}/(\pi a^2)$, of amorphous silicate dust with size $a=\SI{0.1}{\micron}$. The values of the model parameters used in the calculations are shown in Tab.\,\ref{tab:parameters}. The dependence on individual parameters is summarized for each emission and absorption process. Since the analytical approximate solution for the absorption coefficients reproduces the numerical solution well, we do not distinguish between them in this section.
\begin{table*}[t]
    \centering
    \caption{Model parameters in the SP model.}
    \begin{threeparttable}
    \begin{tabular}{lllll}
        \hline \hline
        Variable & Meaning & Unit & Value & Reference \\
        \hline
        \multicolumn{5}{l}{free parameters} \\
        \hline
        $W$ & energy scale of $V_\mathrm{SP}$ & \si{erg} & --- & --- \\
        $D_2^{\max}$ & upper cutoff of $D_2$ & --- & --- & --- \\
        $\gamma$ & phase relaxation rate for Bloch vector & $\si{s^{-1}}$ & --- & --- \\
        $T$ & dust temperature & \si{K} & --- & --- \\
        $f_\mathrm{SP}$ & number ratio of atoms trapped in $V_\mathrm{SP}$ & --- & --- & --- \\
        \hline
        \multicolumn{5}{l}{given parameters} \\
        \hline
        $m$ & atomic mass & g & $M_\mathrm{sil}/7/N_\mathrm{A}$ \tnote{$*$} & \cite{Li+01} \\
        $qx_0$ \tnote{$\dagger$} & atomic electric dipole & \si{D} & 1 & \cite{Bosch78} \\ 
        $a$ & dust radius & \si{\micron} & 0.1 & --- \\
        $\varrho$ & dust mass density & \si{g.cm^{-3}} & 3.5 & \cite{Li+01} \\
        $N_\mathrm{atom}$ & number of atoms composing dust & --- & $\varrho V /m$ & --- \\
        $c_t$ & sound velocity for transverse wave & \si{cm.s^{-1}} & \num{3e5} & \cite{Bosch78} \\
        $\Lambda_t$ & elastic dipole for transverse wave & \si{eV} & 1 & \cite{Meny+07} \\
        $\Gamma_0$ & attempt frequency & \si{s^{-1}} & \num{1e13} & \cite{Bosch78} \\
        $l_c$ & coherent length & \si{nm} & 3 & \cite{Bosch78} \\
        $\braket[1]{q^2/m}$ & coefficient in the DCD model & \si{esu^2.g^{-1}} & $e^2/m_\mathrm{O}$ \tnote{$\ddagger$} & \cite{Meny+07} \\
        \hline
    \end{tabular}
    \begin{tablenotes}\footnotesize
        \item[$*$] The molar mass of amorphous silicate is $M_\mathrm{sil} = \SI{172.2}{g.mol^{-1}}$ 
        when its chemical formula is assumed to be $\mathrm{MgFeSiO_4}$.
        The Avogadro constant is denoted by $N_\mathrm{A}$.
        \item[$\dagger$] The product of the independent parameters $q$ and $x_0$ is treated as a single parameter.
        \item[$\ddagger$] Here $e$ is the electron charge and $m_\mathrm{O}$ is the mass of the oxygen atom.
    \end{tablenotes}
    \label{tab:parameters}
    \end{threeparttable}
\end{table*}

\begin{figure*}[t]
\centering
\includegraphics[scale=1]{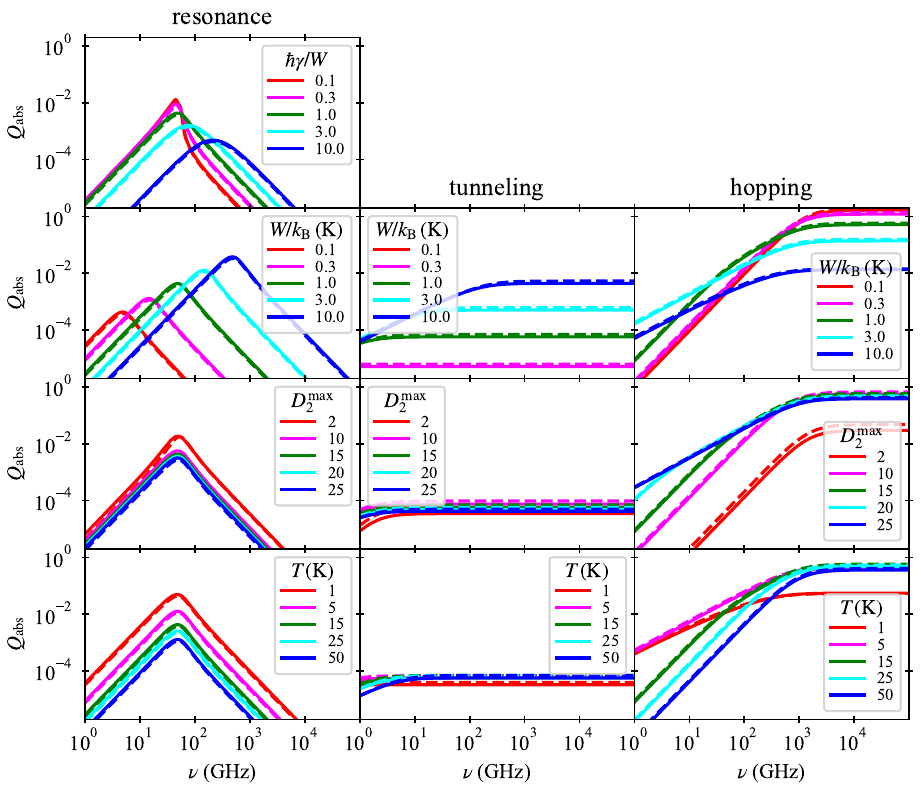}
\caption{
    Frequency dependence of the absorption coefficient of amorphous silicate dust with $a=\SI{0.1}{\micron}$. Solid and dashed curves in each panel represent numerical and analytical solutions, respectively. The left, middle, and right columns represent the contributions from resonance, tunneling, and hopping relaxation processes, respectively. From the top, $\gamma$, $W$, $D_2^\mathrm{max}$, and $T$ are treated as parameters, and the rest are fixed at $\hbar\gamma/W=1,\ W/k_\mathrm{B}=\SI{1}{K}$, $D_2^\mathrm{max}=15$, and $T=\SI{15}{K}$. For all curves, $f_\mathrm{SP} = 1$. Since $\gamma$ does not contribute to the tunneling and hopping relaxation processes, the corresponding panels are omitted.
}
\label{fig:Qabs}
\end{figure*}

\subsubsection{Resonance} 
\label{sec:Qabs_res}
The left column of Fig.\,\ref{fig:Qabs} shows the absorption coefficients due to resonance transitions, $Q_\mathrm{abs}^\mathrm{res}$, which have peaks at frequencies roughly corresponding to the inter-level energy difference ($2\pi\nu \sim \omega_0$).

For the phase relaxation rate $\gamma$, the peak gets sharper as $\gamma$ decreases. This is because as $\gamma$ increases, the resonance transition time becomes shorter, and the fluctuation of the resonance transition frequency gets larger. As $\gamma$ increases, the peak position of the resonance transition shifts toward the larger frequency side and peaks at around $\omega = \gamma$. This is because the imaginary part of the component of the Bloch vector $u_\pm^{(\omega)}$ asymptotically approaches $\mathrm{Im}(u_\pm^{(\omega)}) \propto \gamma/(\gamma^2+\omega^2)$ when $\gamma \gg \omega_0$ (see Eq.\,\eqref{eq:u_pm}).

For the energy scale $W$, the peak position of the resonance transition shifts toward a larger frequency value as $W$ increases. This is a straightforward concept: the larger $W$, the larger $\epsilon$, and consequently, the larger the resonance frequency, $\omega_0$ (see Eq.\,\eqref{eq:u_pm}). The increase in amplitude as $W$ increases is due to the hyperbolic tangent function dependence of the Bloch vector as seen in Eq.\,\eqref{eq:u_pm}.

For $D_2^\mathrm{max}$, which is the maximum value of $D_2$, $Q_\mathrm{abs}^\mathrm{res}$ decreases as $D_2^\mathrm{max}$ increases. This is because the minimum value of $\epsilon$ decreases as $D_2^\mathrm{max}$ increases. The contribution from the larger $\epsilon$ buries the contribution from the smaller $\epsilon$ and thus does not affect the frequency dependence of $Q_\mathrm{abs}^\mathrm{res}$ enough to change it. On the other hand, due to the assumption of uniformity of the distribution function, as $D_2^\mathrm{max}$ increases, the relative contribution from the larger $\epsilon$ to the frequency dependence of $Q_\mathrm{abs}^\mathrm{res}$ decreases, and the amplitude of the overall $Q_\mathrm{abs}^\mathrm{res}$ becomes smaller.

For the dust temperature $T$, $Q_\mathrm{abs}^\mathrm{res}$ decreases as $T$ increases. This is because as $T$ increases and the difference between the occupancy of the ground and excited states decreases, resonance transitions generally occur less frequently.

\subsubsection{Tunneling relaxation} 
\label{sec:Qabs_tun}
The middle column of Fig.\,\ref{fig:Qabs} shows the absorption coefficients due to tunneling relaxation, $Q_\mathrm{abs}^\mathrm{tun}$. At low frequencies ($\omega \ll \Gamma_\mathrm{tun}$), $Q_\mathrm{abs}^\mathrm{tun}$ increases with increasing frequency but remain constant at $\omega \gtrsim \Gamma_\mathrm{tun}$. It is clear from the asymptotic form $\mathrm{Im}(w^{(\omega)}) \propto \Gamma/\omega$ in the high-frequency limit (see Eq.\,\eqref{eq:w}). It is also qualitatively understood from the fact that a tunneling effect occurs with a certain probability when a photon with energy greater than $\hbar\Gamma_\mathrm{tun}$ is injected.

As $W$ increases, the frequency at which $Q_\mathrm{abs}^\mathrm{tun}$ switches to constant increases. This is because $\Gamma_\mathrm{tun}$, corresponding to the switching frequency, increases monotonically with $W$ (see Eq.\,\eqref{eq:Gamma-tun}).

For $D_2^\mathrm{max}$, $Q_\mathrm{abs}^\mathrm{tun}$ decreases as $D_2^\mathrm{max}$ increases. This is for the same reason as in the case of resonance transitions. However, for $D_2^\mathrm{max} = 2$, $Q_\mathrm{abs}^\mathrm{tun}$ becomes smaller. This is because as $D_2^\mathrm{max}$ decreases, the potential barrier becomes lower in most regions of the $D_1\text{--}D_2$ plane, making the tunneling effect less likely.

For $T$, the switching frequency and amplitude of $Q_\mathrm{abs}^\mathrm{tun}$ change monotonically. This is because it depends on the balance between the temperature dependence of $\Gamma_\mathrm{tun}$ and $w^{(\omega)}_\mathrm{tun}$. Since $\Gamma_\mathrm{tun}$ increases consistently with $T$, the higher $T$, the better the relaxation efficiency, and the greater the amplitude at lower frequencies. Note that, however, the temperature dependence is insignificant in the sub-mm.

\subsubsection{Hopping relaxation} 
\label{sec:Qabs_hop}
The right column of Fig.\,\ref{fig:Qabs} shows the absorption coefficients due to the hopping relaxation, $Q_\mathrm{abs}^\mathrm{hop}$. Similar to the tunneling relaxation, $Q_\mathrm{abs}^\mathrm{hop}$ increases with increasing frequency at the low-frequency side ($\omega \ll \Gamma_\mathrm{hop}$), but at the high-frequency side ($\omega \gtrsim \Gamma_\mathrm{hop}$), $Q_\mathrm{abs}^\mathrm{hop}$ is constant. On the low-frequency side, unlike tunneling relaxation, the spectral index has a value between one and two. This difference in frequency dependence on the low-frequency side comes from the relationship between the relaxation rate and frequency. Since no relaxation occurs on timescales longer than the lower limit of $\Gamma_\mathrm{hop}$, $\Gamma_\mathrm{hop}^\mathrm{min}$, $Q_\mathrm{abs}^\mathrm{hop} \propto \nu^2$ from the low-frequency limit ($\omega \ll \Gamma_\mathrm{hop}^\mathrm{min}$), as seen in Eq.\,\eqref{eq:w}. On the other hand, for $\omega > \Gamma_\mathrm{hop}^\mathrm{min}$, the envelope of absorption coefficients from atoms with various $\Gamma_\mathrm{hop}$ values determines the frequency dependence of the total $Q_\mathrm{abs}^\mathrm{hop}$, indicating $Q_\mathrm{abs}^\mathrm{hop} \propto \nu$. This feature is similar in the tunneling relaxation but appears at much lower frequencies, which is not shown in Fig.\,\ref{fig:Qabs} because $\Gamma_\mathrm{tun}$ is too small.

For $W$, the amplitude at higher frequencies increases with increasing $W$, and the frequency dependence at lower frequencies becomes flatter, which is the opposite of tunneling relaxation. This is because $\Gamma_\mathrm{hop}$ decreases monotonically for $W$ (see Eq.\,\eqref{eq:Gamma-hop}).

For $D_2^\mathrm{max}$, the maximum value of $V_b$ increases as $D_2^\mathrm{max}$ increases (see Eq.\,\eqref{eq:Vb_approx}), and as $\Gamma_\mathrm{hop}^\mathrm{min}$ decreases, the frequency dependence on the low-frequency side changes.

For $T$, for the same value of $V_b$, the higher $T$, the smaller $\Gamma_\mathrm{hop}$ becomes (see Eq.\,\eqref{eq:Gamma-hop}), so the frequency dependence depends on the relationship between the relaxation rate and frequency as with other parameters.

\subsection{Disordered charge distribution model}
\label{sec:DCD}
Thermal emission from amorphous dust is affected not only by radiation processes related to the atomic TLS but also by contributions from lattice vibrations \citep{Meny+07}. As an electric interaction in amorphous dust, we considered photon absorption by acoustic vibrations in a disordered charge distribution (DCD) proposed by \cite{Schlomann64}. The electric polarizability in the DCD model is given as \citep{Meny+07, Nashimoto+20pasj}, 
\begin{align}
    \mathrm{Im}\pab{\alpha_\mathrm{DCD}} =
    \frac{V\omega}{12\pi c_t^3} \braket[1]{\frac{q^2}{m}}
    \bab{1-\pab{1+\pab{\frac{\omega}{\omega_c}}^2}^{-2}},
    \label{eq:alp_dcd}
\end{align}
where $\omega_c \equiv c_t/l_c$, and $l_c$ is the coherent length characterizing the irregular charge distribution and is treated as a fixed value (see Tab.\,\ref{tab:parameters}). \cite{Paradis+11} insisted that $l_c$ is a critical parameter in SED fitting for observational data, but for simplicity, it is treated as a given parameter in this paper, where the impact of fixing the parameter value is discussed in Sec.\,\ref{sec:Impact_SP}.

The term due to transverse sound waves is ignored because $c_l^{-3} \ll c_t^{-3}$ \citep{Ramos+97}. The parameter $\braket[1]{q^2/m}$, which represents the average value of the ratio of the square of the charge of the particles to their mass, which carry lattice vibration, uses the same value as \cite{Meny+07}. By substituting Eq.\,\eqref{eq:alp_dcd} into Eqs.\,\eqref{eq:Clausius-Mossotti} and \eqref{eq:Cabs}, the complex permittivity and absorption cross-section derived from the DCD model can be calculated, respectively.
\section{Comparison with Observations} 
\label{sec:CompObs}

We compared the Perseus molecular cloud (MC) spectrum with our amorphous dust emission model. Model fitting was performed the same way as for \cite{Nashimoto+20pasj} using the observation data provided by \cite{QUIJOTE1}. Here is a brief explanation of the data to be compared; for details, see \cite{QUIJOTE1}. We used the observation data given by \cite{Haslam+82}, \cite{Berkhuijsen72}, \cite{Reich+86}, COSMOSOMAS \citep{Watson+05}, QUIJOTE \citep{QUIJOTE1}, WMAP \citep{Bennett+13}, Planck \citep{PlanckI+14} and COBE/DIRBE \citep{Hauser+98}. The maps, which originally had different resolutions, were convolved to match the resolution to \ang{1}. Although the contribution of CO has been subtracted, the Planck \SIlist{100;217}{GHz} data may still contain CO contamination, so they were not used in this study. By integrating within a \ang{1.7} radius of the Perseus MC and subtracting the data from the outer ring, the foreground and background for the MC were removed.

Since radiation is equivalent to absorption from Kirchhoff's law (c.f., \citealp{Rybicki&Lightman}), the emission spectrum of amorphous dust, $I_\nu^d$, is expressed using the absorption cross-section as,
\begin{align}
    I_\nu^d = N_d C_\mathrm{abs} B_\nu(T),
\end{align}
where $N_d$ is the dust column density and $B_\nu(T)$ is the Planck function. Given that the contribution from large dust grains is dominant at longer wavelengths than in the FIR, we do not factor in the dust size distribution and assume a typical size of $a = \SI{0.1}{\micron}$. We also assume that the dust is in thermal equilibrium with the ISRF and do not consider the temperature distribution. In reality, interstellar dust has a size distribution and different equilibrium temperatures depending on the size, but when the temperature range is only a few K, the effect on the overall shape of the spectrum may be small enough (see Fig.\,\ref{fig:Qabs}). Therefore, this assumption is reasonable, except that it does not account for the effects of stochastic heating on nanoscale dust particles.

\begin{figure*}[!t]
\includegraphics[scale=1]{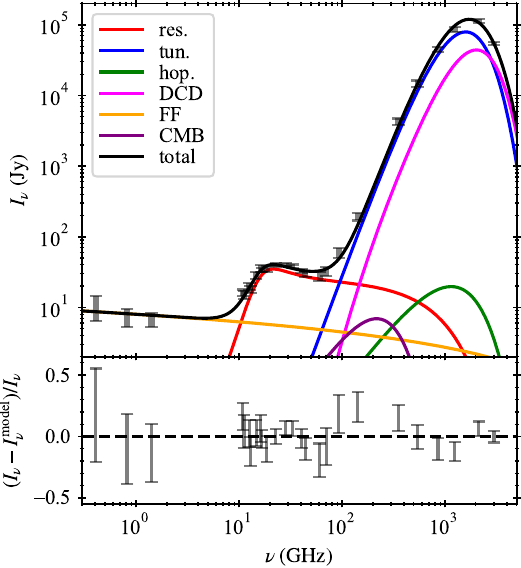}
\includegraphics[scale=1]{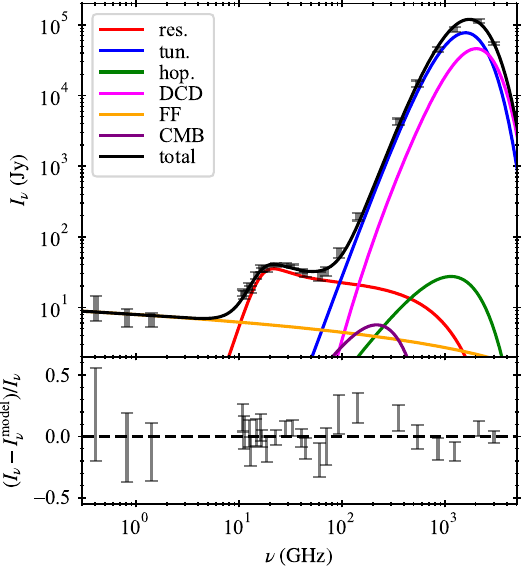}
\caption{
    Observed spectra and best-fit models for the Perseus MC. The left and right panels show the best-fit analytical and numerical solution models, respectively. Data points are given by table 2 in \cite{QUIJOTE1}. The red, blue, green, and magenta curves represent the contributions of resonance transition, tunnel relaxation, hopping relaxation, and DCD in dust emission, respectively. The orange curve is the free-free emission, the purple curve is the CMB temperature fluctuation, and the black curve is the total. The CMB spectrum is taken in absolute value. The lower panels show the differences from the best-fit models.
}
\label{fig:SED_sp}
\end{figure*}
\begin{table}[!t]
    \centering
    \caption{Best-fit parameter values for the Perseus MC.}
    \begin{tabular}{cr@{}lr@{$\pm$}lc}
        \hline \hline
        Variable & \multicolumn{4}{c}{Value} & Unit \\
        \hline
        & \multicolumn{2}{c}{analytical} & \multicolumn{2}{c}{numerical} \\
        \hline
        $W$ & 3.02 & $\pm$0.03 & 3.29 & 0.04 &
        $\SI{e-1}{K} \times k_\mathrm{B}$ \\
        $D_2^{\max}$ & 1.65 & $^{+0.01}_{-0.00}$ & 2.11 & 0.02 &
        --- \\
        $\gamma$ & 4.81 & $^{+0.18}_{-0.17}$ & 4.72 & 0.18 &
        \SI{e10}{s^{-1}}  \\
        $T$ & 19.6 & $\pm$0.1 & 19.5 & 0.1 &
        \si{K}  \\
        $f_\mathrm{SP}$ & 95.1 & $\pm$2.2 & 59.4 & 1.4 &
        \% \\
        $N_d$  & 7.68 & $\pm$0.15 & 8.12 & 0.16 &
        \SI{e7}{cm^{-2}} \\
        EM & 26.7 & $\pm$2.4 & 26.4 & 2.4 &
        \si{cm^{-6}.pc} \\
        $\varDelta T_\mathrm{CMB}$ & 
        $-5.17$ & $\pm$6.10 & $-4.23$ & 6.10 &
        \si{\mu K} \\
        \hline
        $\chi^2_\mathrm{red}$ & \multicolumn{2}{c}{1.33} & \multicolumn{2}{c}{1.29} & --- \\
        \hline
    \end{tabular}
    \label{tab:best-fit}
\end{table}

Fig.\,\ref{fig:SED_sp} shows the observed spectrum and the best-fit models for the Perseus MC, and Tab.\,\ref{tab:best-fit} shows their parameter values. The free-free emission and CMB temperature fluctuation spectra are calculated using the formula in \cite{Planck+11}, where the electron temperature and the mean CMB temperature are set to $T_e = \SI{8000}{K}$ and $T_\mathrm{CMB} = \SI{2.725}{K}$, respectively. The emission measure, EM, and the CMB temperature fluctuation, $\varDelta T_\mathrm{CMB}$, are treated as fitting parameters.
Synchrotron radiation, a component of interstellar radiation, is assumed to be negligible. We performed the fitting using the Levenberg–Marquardt algorithm to minimize the chi-square statistic. The uncertainty of each fitted parameter was estimated by evaluating the parameter ranges over which the chi-square value increases by one from its minimum value. This approach corresponds to the $1\sigma$ confidence intervals under the assumption of Gaussian-distributed residuals and locally linear response of the model with respect to the parameters. In the SED fitting conducted in this study, we confirmed that the solution avoided dependence on initial values and convergence to local minima, and successfully reached the global optimum.

The left panel in Fig.\,\ref{fig:SED_sp} shows that the amorphous dust emission based on the SP model calculated using the analytical approximate solution (see Sec.\,\ref{sec:SP_ana}) can accurately reproduce the observed data from FIR to microwave. The reduced chi-square value for the best-fit model is $\chi_\mathrm{red}^2 = 1.33$ (the degree of freedom is 18). The reduced chi-square value for the same data fitted by the TLS model is $\chi_\mathrm{red}^2 = 1.67$ \citep{Nashimoto+20pasj}. Judging by the reduced chi-square alone, the SP model can reproduce the observed data more accurately than the TLS model, although only slightly. The tunnel relaxation dominates in the frequency range from \SI{100}{GHz} to \SI{1}{THz}, while the contribution of the resonance transition dominates in the frequency range from \SI{10}{GHz} to \SI{100}{GHz}. The contribution of the thermal relaxation is very small, unlike in the TLS model \citep{Paradis+11}. In the low-frequency region below the FIR, the contribution of the DCD is not significant, so fixing the correlation length, $l_c$, that determines the frequency at which the spectral index of the absorption cross-section changes from 2 to 4 does not significantly affect the fitting accuracy.

\begin{figure}[t]
\centering
\includegraphics[scale=1]{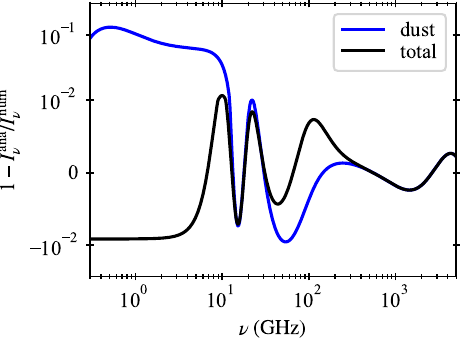}
\caption{
    Difference between the best-fit analytical and numerical solution models for the Perseus MC. The blue and black curves show the dust emission and the total SED, adding the free-free emission and CMB, respectively.
}
\label{fig:SED_sp_diff}
\end{figure}
While the amorphous dust emission based on the SP model with analytical approximate solutions can reasonably reproduce the observed spectrum of the Perseus MC, the parameter value for $D_2^\mathrm{max}$ estimated by the best-fit model may not be applicable to the STA (see Sec.\,\ref{sec:SP_ana}). The right panel of Fig.\,\ref{fig:SED_sp} shows the result of model fitting using a numerically computed SP model for similar data. Like the analytical approximation, the numerical solution can also reproduce the observed data with outstanding accuracy. The reduced chi-square value for the numerical solution is $\chi^2_\mathrm{red}=1.29$ and both solutions have no significant difference in the reduced chi-square. The best-fit parameter values of both solutions for $W$, $D_2^\mathrm{max}$, $f_\mathrm{SP}$, and $N_d$ differ beyond the error range, as presented in Tab.\,\ref{tab:best-fit}. Although our estimations of $T$ seem a bit high for MCs shown in Tab.\,\ref{tab:best-fit}, the values are consistent with previous studies using other dust models \citep{Schnee+08, Planck+11, QUIJOTE1}, and this result is not specific to the SP and/or DCD models. Fig.\,\ref{fig:SED_sp_diff} shows the difference between the analytical and numerical best-fit models, and both the dust and total SEDs show a difference of less than 1\% in the frequency bands higher than \SI{10}{GHz}. The properties of amorphous dust expected from both best-fit models are discussed in Sec.\,\ref{sec:params}.
\section{Discussion} 
\label{sec:discussion}

\subsection{Comparison with the TLS model}
A comparison of the SP and TLS models will be discussed. The conversion of the physical parameters describing both models allows a fair comparison. Absorption coefficients calculated under equivalent conditions are compared to clarify the differences between both models.

\subsubsection{Conversion from the TLS model to the SP model}
In the TLS model, the distribution function for $\Delta$ and $\Delta_0$, $f(\Delta,\Delta_0)$, is given by \cite{Phillips72} as,
\begin{align}
    f(\Delta, \Delta_0) = \frac{P_0}{\Delta_0},
    \label{eq:dist_Del_Del0}
\end{align}
where $P_0$ is the normalization constant of $f(\Delta, \Delta_0)$ and is obtained as,
\begin{align}
    P_0 =&
    \pab{ \int_{\Delta_0^\mathrm{min}}^{\Delta_0^\mathrm{max}} 
    \frac{\odif{\Delta_0}}{\Delta_0}
    \int_0^{\sqrt{\epsilon_\mathrm{max}^2-\Delta_0^2}} \odif{\Delta}}^{-1} .
    \label{eq:P0}
\end{align}
This integral has an elementary analytical solution but is omitted due to its length.
We assume a normal distribution as the distribution function of the potential barrier height $V_b$ \citep{Bosch78},
\begin{align}
    f(V_b) &=
    \begin{cases}
    P_b \exp \pab{-\pab{\frac{V_b-V_{b,m}}{V_{b,0}}}^2}
    & V_b \geq V_{b,\mathrm{min}}, \\
    0 & V_b < V_{b,\mathrm{min}},
    \end{cases}
    \label{eq:f_Vb_tls} \\
    P_b &= 
    \frac{2}{V_{b,0}\sqrt{\pi}}
    \pab{\mathrm{Erf}\pab{\tfrac{V_{b,m}-V_{b,\mathrm{min}}}{V_{b,0}}}+1},
\end{align}
where $\mathrm{Erf}$ is the error function. We set the parameter values as $V_{b,0}/k_\mathrm{B} = \SI{410}{K}$, $V_{b,m}/k_\mathrm{B} = \SI{550}{K}$, and $V_{b,\mathrm{min}}/k_\mathrm{B} = \SI{50}{K}$ provided by \cite{Bosch78}.

By integrating with these distribution functions, the sum of the polarizability based on the TLS model, $\alpha_\mathrm{TLS}$, is calculated using each atomic polarizability, $\alpha_\mathrm{SP}^i$, as,
\begin{align}
    \frac{\alpha_\mathrm{TLS}}{N_\mathrm{DWP}} =
    \!\iiint\! \odif{\Delta}  \odif{\Delta_0} \odif{V_b}
    f(\Delta,\Delta_0) f(V_b) 
    \alpha_\mathrm{SP}^i(\Delta,\Delta_0,V_b),
\end{align}
where $\alpha_\mathrm{SP}^i$ is given in Eq.\,\eqref{eq:polarizability}.

To compare the SP model with the TLS model, the variables in the TLS model, $\Delta_0^\mathrm{min}$ and $\Delta_0^\mathrm{max}$, are corresponded to those in the SP model, $W$ and $D_2^\mathrm{max}$. From Eq.\,\eqref{eq:Delta0_approx}, $\Delta_0^\mathrm{min}$ and $\Delta_0^\mathrm{max}$ are provided as,
\begin{align}
    \Delta_0^\mathrm{min} &=
    W \pab{D_2^\mathrm{max}}^{3/2}
    \exp \pab{1-\frac{\sqrt{2}}{3}\pab{D_2^\mathrm{max}}^{3/2}},
    \label{eq:Delta0_min} \\
    \Delta_0^\mathrm{max} &= \frac{3W}{\sqrt{2}}.
    \label{eq:Delta0_max}
\end{align}
In both models, $T$, $\gamma$, and $f_\mathrm{SP}$ have the same meaning.

\begin{figure}[t]
\centering
\includegraphics[scale=1]{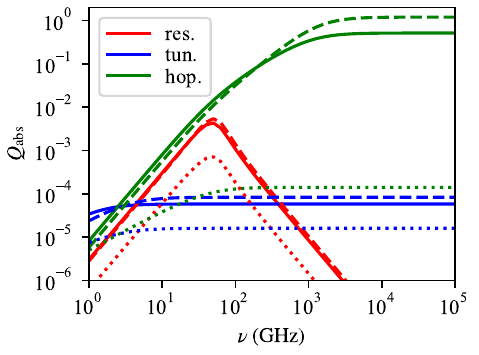}
\caption{
    Frequency dependence of the absorption coefficients of amorphous silicate dust with $a=\SI{0.1}{\micron}$. We set $W/k_\mathrm{B} = \SI{1}{K}$, $D_2^\mathrm{max} = 15$, $T = \SI{15}{K}$, $\gamma = W/\hbar$, and $f_\mathrm{SP} = 1$. The colors of the curves indicate differences in the emission mechanism. The solid and dotted curves show the results based on the SP and TLS models, respectively. The dashed curves show the TLS model results when revising the distribution function (see Sec.\,\ref{sec:rev-TLS}).
}
\label{fig:Qabs_tls}
\end{figure}
Fig.\,\ref{fig:Qabs_tls} shows the absorption coefficients calculated using the SP and TLS models. In each model, we set $W/k_\mathrm{B} = \SI{1}{K}$, $D_2^\mathrm{max} = 15$, $T = \SI{15}{K}$, $\gamma = W/\hbar$, and $f_\mathrm{SP} = 1$. The absorption coefficients for resonance transitions and tunneling relaxation are consistent with the frequency dependence in both models. However, the amplitude tends to be smaller in the TLS model. The absorption coefficients for hopping relaxation, on the other hand, show different results for the two models. This is mainly due to the different distribution ranges of the potential barrier $V_b$ in the two models. From Eq.\,\eqref{eq:Vb_approx}, $0 \leq V_b \leq W(D_2^\mathrm{max}/2)^2$ in the SP model. Substituting $W/k_\mathrm{B} = \SI{1}{K}$ and $D_2^\mathrm{max} = 15$, the maximum value of $V_b$ is about $\SI{56}{K} \times k_\mathrm{B}$, an order of magnitude smaller than $V_{b,m}$ in the TLS model. Given the typical values of $W$ and $D_2^\mathrm{max}$ estimated from laboratory measurements (see Sec.\,\ref{sec:params}), the SP model is unlikely to consistently reproduce the potential barrier values expected from the TLS model.

\subsubsection{Revising the TLS model} 
\label{sec:rev-TLS}
We constructed a correction method for the TLS model that asymptotically approaches the results of the SP model. From $f(D_1,D_2)\odif{D_1}\odif{D_2} = f(\Delta,\Delta_0)\odif{\Delta}\odif{\Delta_0}$, then $P_\mathrm{SP}$ and $P_0$ have the following relationship,
\begin{align}
    P_0 = 
    \frac{P_\mathrm{SP}}
    {W\sqrt{D_2-1}\pab{\tfrac{3}{\sqrt{2}D_2}-\sqrt{D_2}}}.
    \label{eq:P0_sp}
\end{align}
To write $P_0$ as a function of $\Delta_0$, an inverse function of Eq.\,\eqref{eq:Delta0_approx} is introduced as,
\begin{align}
    D_2 &= 
    \pab{\frac{9}{2}}^{1/3} 
    W_\mathrm{L}\pab{-\frac{\sqrt{2}}{3e} \frac{\Delta_0}{W}}^{2/3},
    \label{eq:D2}
\end{align}
where $e$ is Napier's constant, and $W_\mathrm{L}$ is the Lambert W function satisfying $W_\mathrm{L}(x)\exp(W_\mathrm{L}(x)) = x$.

For the distribution function of the potential barrier, $f(V_b)$, $V_b$ can be regarded as a function of $D_2$ because of Eq.\,\eqref{eq:Vb_approx}, so $f(V_b)$ is obtained as,
\begin{align}
    f(V_b) = 
    \odv{D_2}{V_b}f(D_2) = 
    \frac{2}{WD_2} f(D_2),
    \label{eq:f_Vb_D2}
\end{align}
where $f(D_2)$ is given by,
\begin{align}
    f(D_2) =
    2\int \odif{D_1} f(D_1,D_2).
    \label{eq:f_D2}
\end{align}
where the coefficient 2 comes from the even symmetry of the $D_1$ distribution.

Fig.\,\ref{fig:Qabs_tls} also shows the frequency dependence of the absorption coefficients calculated using Eqs.\,\eqref{eq:P0_sp} and \eqref{eq:f_Vb_D2} instead of Eqs.\,\eqref{eq:P0} and \eqref{eq:f_Vb_tls}. The absorption coefficients derived from resonance transitions and tunneling relaxation are roughly in agreement. On the other hand, the absorption coefficients due to hopping relaxation are consistent at low frequencies, but the amplitude differs by a factor of 2 at high frequencies. This may be because $V_b$ is regarded as an independent variable in the TLS model. We conclude that the TLS model cannot reproduce the absorption coefficient expected from the SP model by simply correcting the distribution function of the parameters, and the SP model must be calculated to estimate the amorphous dust emission.

\subsection{Comparison of the best-fit models} 
\label{sec:params}
\begin{table*}[t]
    \centering
    \caption{Comparison of physical quantities of amorphous dust in the best-fit models.}
    \begin{threeparttable}
    \begin{tabular}{cccccc}
        \hline \hline
        & $W$ & $\Delta_0^\mathrm{max}$ & $\Delta_0^\mathrm{min}$ & $\gamma$ & $N_\mathrm{DWP}/(\varrho V)$ \\
        & ($\si{K} \times k_\mathrm{B}$) & ($\si{K} \times k_\mathrm{B}$) 
        & ($\si{K} \times k_\mathrm{B}$) & (\SI{e10}{s^{-1}}) 
        & (\si{g^{-1}}) \\
        \hline
        analytical &
        0.302 & 0.641 & 0.641 & 4.81 & \num{2.33e22} \\
        numerical & 
        0.329 & 0.698 & 0.646 & 4.72 & \num{1.45e22} \\
        reference &
        1--5\tnote{$*$} & 20.6\tnote{$\dagger$} & $\lesssim \num{2e-3}$\tnote{$\dagger$} & 0.3\tnote{$\ddagger$} & 
        \num{e16}--\num{e17}\tnote{$*$} \\
        \hline
    \end{tabular}
    \begin{tablenotes} \footnotesize
        \item[$*$] \cite{Ramos+97}.
        \item[$\dagger$] \cite{Meny+07}.
        \item[$\ddagger$] \cite{VonSchickfus&Hunklinger77}.
    \end{tablenotes}
    \end{threeparttable}
    \label{tab:params_est}
\end{table*}
First, regarding the comparison between the numerical and analytical models, only $f_\mathrm{SP}$ shows a significant difference among the physical properties of the amorphous dust estimated by both models. For $f_\mathrm{SP}$, the analytical approximation requires about 1.6 times more atoms per unit mass to be bound to the DWP than the numerical solution. The SP model (and also the TLS model) assumes that amorphous deformation of the crystalline occurs when the crystal lattice is distorted, and the atoms contributing to the distortion are trapped in the DWP. Therefore, as the number of such trapped atoms increases, the degree of disorder in the crystal structure of the entire material also increases, so $f_\mathrm{SP}$ is an indicator of the internal structure of the amorphous dust.

Although the analytical approximation can reproduce the observed data with good accuracy, a numerical solution is necessary to extract information from the dust accurately. However, this result does not mean that the analytical approximate solution is not inapplicable to amorphous dust. We do not rule out the possibility that the analytical solution could not be applied because the Perseus MC happened to be an object surrounded by amorphous dust with a small $D_2^\mathrm{max}$. Increasing the number of samples will verify the practicality of the analytical solution.

Then, regarding the comparison with laboratory measurement, Tab.\,\ref{tab:params_est} displays the parameter values of the best-fit models obtained through SED fitting to the Perseus MC and compares them with typical values estimated from laboratory measurements of amorphous materials.
For $W$, it is known that the value typically estimated from measurements of the crossing temperature of the temperature dependence of the heat capacity is $W/k_\mathrm{B} \simeq \num{1}\text{--}\SI{5}{K}$ (e.g., see table 1 in \citealp{Ramos+97}). The best-fit models are one order of magnitude smaller than this value. \cite{Meny+07} adopted $\Delta_0^\mathrm{max}/k_\mathrm{B} = \SI{20.6}{K}$ as the typical value in the TLS model based on laboratory spectroscopic measurements of amorphous silicate. The best-fit values are about $\Delta_0^\mathrm{max}/k_\mathrm{B} \simeq \SI{0.6}{K}$, which are also more than one order of magnitude smaller than the reference value. For $\Delta_0^\mathrm{min}$, the best-fit values calculated by Eq.\,\eqref{eq:Delta0_min} are $\Delta_0^\mathrm{min}/k_\mathrm{B} \simeq \SI{0.6}{K}$. The temperature dependence of the low-temperature heat capacity of amorphous silicate follows the TLS model at least up to $T = \SI{2}{mK}$, which is estimated to be $\Delta_0^\mathrm{min}/k_\mathrm{B} \lesssim \SI{2}{mK}$ \citep{Meny+07}. The best-fit model predicts a value of $\Delta_0^\mathrm{min}$ at least two orders of magnitude larger than the results of laboratory measurements. The phase relaxation rate $\gamma$ was measured as $\gamma = \SI{3e9}{s^{-1}}$ for vitreous silica \citep{VonSchickfus&Hunklinger77}. Our best-fit model value is about one order of magnitude larger than this value. Compared to the ratio to the energy scale $W$, the best-fit model is $\hbar\gamma/W \sim 1$, while the value estimated from laboratory measurements is $\hbar\gamma/W \sim \num{e-2}$. In other words, the best-fit model is two orders of magnitude larger than the laboratory measurement. Laboratory measurements of low-temperature heat capacity have estimated the number of atoms trapped in the DWP per unit mass of the material to be \num{e16}--\SI{e17}{g^{-1}} \citep{Ramos+97}. Using the best-fit value of $f_\mathrm{SP}$, the value is calculated to be $N_\mathrm{DWP}/(\varrho V) \sim$ \SI{e22}{g^{-1}}. The best-fit model requires 5--6 orders of magnitude more atoms to be trapped in $V_\mathrm{SP}$ than typical values estimated from laboratory measurements.

If certain parameters are anti-correlated, a change in one could cause a significant change in another. This could reduce some of the discrepancies observed in relation to the laboratory data shown in Tab.\,\ref{tab:params_est}. However, the error bars of the model parameters are small compared to the discrepancies observed when comparing with the experimental data (see Tab.\,\ref{tab:best-fit}), so these discrepancies could not be attributed solely to the correlation or inverse correlation between the model parameters.

The reference values in Tab.\,\ref{tab:params_est} represent typical physical properties of amorphous materials under various compositions and different measurement conditions and do not correspond to the physical properties of interstellar dust, such as the microstructure, porosity, shape, and composition of dust particles. Note that, therefore, these values are not directly comparable to observational results of interstellar dust.

The discrepancies between the observational results and the reference measurements shown in Tab.\,\ref{tab:params_est} should not necessarily be interpreted as evidence for substantial differences in the amorphous properties of materials in interstellar space versus those on Earth. Rather, they highlight ongoing gaps in our observational, theoretical, and experimental understanding of amorphous dust. To address these gaps, it is important to (1) conduct laboratory measurements using analog materials that simulate interstellar dust, (2) investigate the relationship between model parameters and the microscopic structure and composition of dust, and (3) estimate model parameters by fitting observational data with models that include multiple dust components. These remain important challenges for future studies aiming to refine our understanding of the physical properties of interstellar dust.

\subsection{Impact of applying the SP model}
\label{sec:Impact_SP}
We discuss the impacts of applying our amorphous dust emission model based on the SP model to astronomical observations.

Even if the SP model is applied instead of the TLS model, it can only reproduce the observation data for the Perseus MC with almost the same level of accuracy as shown in Sec.\,\ref{sec:CompObs}. However, the SP model suggests that the amorphous dust comprises atoms trapped in the DWP with a lower potential barrier than the TLS model. The number of atoms that can thermally jump over the potential barrier is large, and the emission intensity due to the hopping relaxation is relatively large. In other words, while maintaining the same AME (resonant transition emission) intensity in the SP model and TLS model, the SP model can increase the radiation intensity in the sub-mm waveband originating from hopping relaxation and achieve a smaller $\beta$. The spectral index of the Perseus MC, which was the target of the SED fitting in this paper, was estimated to be $\beta = 1.73 \pm 0.11$ by \cite{QUIJOTE1}, while the typical value for the Galactic halo region is $\beta = 1.62$ \citep{PlanckXI+14}. Both estimations used a similar dataset as this study, with $\beta$ determined using data at wavelengths longer than \SI{100}{\micro m}. The reason why the TLS model was able to reproduce the SED of the Perseus MC with good accuracy may be that the Perseus MC has a relatively large $\beta$. It may be possible that the SP model can explain regions with low $\beta$ (as well as with significant AME intensity) where the TLS model cannot, and it is crucial to verify the SP model using such objects and regions. Of course, to evaluate whether the origin of AME is large amorphous grains or not, comparison with the spinning dust model \citep{Draine&Lazarian98feb, Draine&Lazarian98nov} is important. Since both models can reproduce the dust SED well, it is necessary to capture features or trends that are predicted only by one of the models. Resonance transition emission increases in intensity with temperature, so investigating whether there is a positive correlation between AME intensity and dust temperature would be one verification method if the AME originates from large amorphous grains. To explore the origin of the AME, it is essential to analyze the correlation between the AME and the intensity of various dust species. Observations concerning the relationship between the AME and the PAH, the most likely dust component as a carrier in the spinning dust model, vary across different regions and remain unclear (e.g., \citealp{Tibbs+11, Tibbs+12, Hensley+16, Hensley+22, Planck+16, Greaves+18, Bell+19, ArceTord+20}). Besides PAHs, other carrier candidates of the AME include nanosilicates \citep{Hoang+16, Hensley+17, MaciaEscatllar+20}, nanodiamonds \citep{Greaves+18}, and Fe nanoparticles \citep{Hoang&Lazarian16, Hensley+17}. To assess the potential contribution from amorphous large particles, it is necessary to determine if there is a stronger correlation between the AME intensity and the amount of large amorphous grains estimated from the SP model compared to PAHs and other nanoparticles, which is the focus of further work. Even if the origin of AME is spinning dust rather than large amorphous grains, the value of the SP model itself is not lost. This information can be used to constrain the values of parameters related to resonance transitions in the SP model. Narrowing down the range of parameters through observation and measurement in this way can lead to identifying the composition and structure of large amorphous grains.

For the CMB temperature fluctuation, the best-fitting values for the Perseus MC using the SP model are $\varDelta T_\mathrm{CMB} = -5.17\pm6.10 \, \si{\mu K}$ and $\varDelta T_\mathrm{CMB} = -4.23\pm6.10\, \si{\mu K}$ in the analytical and numerical solutions, respectively (see Tab.\,\ref{tab:best-fit}). In previous works, the estimated values are $\varDelta T_\mathrm{CMB} = -19.3^{+6.3}_{-5.9}\, \si{\mu K}$ in the TLS model \citep{Nashimoto+20pasj} and $\varDelta T_\mathrm{CMB} = 22.6\pm13.6\, \si{\mu K}$ using the spinning dust model \citep{QUIJOTE1}, which is the one of the candidates for the origin of the AME \citep{Draine&Lazarian98feb, Draine&Lazarian98nov}. The estimated values of CMB temperature fluctuations differ significantly when assuming the TLS and spinning dust models, but the SP model gives an intermediate value between the two. In observations of the direction of the MCs, since the contribution from dust is significant, it is fundamentally challenging to measure CMB temperature fluctuations with high accuracy. Therefore, our results are not expected to immediately impact the results of cosmological parameter measurements using CMB temperature fluctuations. However, further verification is needed in the future, as the choice of dust model can significantly affect the estimation of CMB temperature fluctuations even in diffuse regions where dust has little effect.

Finally, we discuss the DCD model, which describes the lattice vibrations of amorphous materials. In our model, the value of $\beta$ is determined by the balance between contributions from lattice vibrations in the DCD and relaxation transitions from the SP model. In fact, it is known that the values of $\beta$ have variation in interstellar space (e.g., \citealp{PlanckXI+14}), and a deeper understanding of both models may help resolve this issue. In principle, if the potential function is defined and atomic configurations are known, the DCD and SP models should be solvable in a self-consistent way, and both models are not entirely independent. Therefore, the parameters of the two models are likely to be interrelated, and whether approximations such as fixing DCD parameters as in this paper or treating them as completely free parameters as in \cite{Paradis+11} are valid requires further discussion. Although this is beyond the scope of this paper, it is important to establish a theoretical framework, develop calculation methods, and compare results with experimental data and numerical simulations in order to treat the DCD model and SP model in the same manner.
\section{Conclusion} 
\label{sec:conclusion}
In this paper, we have modeled the optical properties of dust based on the SP model that describes amorphous thermal properties and have revised the amorphous dust thermal emission model in the FIR to microwave wavelengths. Assuming that some atoms composing amorphous dust are trapped in the DWP expressed by a fourth-order function, the interaction between the atomic TLS and the electric field can be directly solved, and the absorption cross-section calculated. In this respect, our proposed amorphous dust emission model is closer to a first-principles model than the conventional TLS model.

The main scientific results and findings of this paper are as follows:
\begin{enumerate}
    \setlength{\parskip}{0cm}
    \setlength{\itemsep}{0cm}
    \item 
    The absorption cross-sections of amorphous dust were calculated numerically and analytically based on the SP model (see Secs.\,\ref{sec:SPmodel} and \ref{sec:Cabs}). When the potential barrier of the DWP was sufficiently high, both results were agreement solutions (see Fig.\,\ref{fig:Qabs}).
    \item 
    When the SP model was compared with the observed data of the Perseus MC from FIR to microwave, we found that it could reproduce the characteristics of the dust long-wavelength emission (see Fig.\,\ref{fig:SED_sp}). The best-fit model has a small value for $D_2^\mathrm{max}$, suggesting that the amorphous dust is composed of atoms trapped in the DWP with a low potential barrier and is thought to be outside the scope of the SP model's analytical approximate solution method (see Tab.\,\ref{tab:best-fit}).
    \item 
    When the SP model and its precursor, the TLS model, were compared under the same conditions, the absorption cross-sections expected from both models were significantly different (see Fig.\,\ref{fig:Qabs_tls}). The difference between the two models is caused by the TLS model treating the energy difference between two levels, $\epsilon$, and the potential barrier, $V_b$, as independent variables, in contrast, they are dependent ones in the SP model. We have also established a conversion of the TLS model to the SP model. However, the corrected TLS model is closer to the SP model but does not match it exactly (see Fig.\,\ref{fig:Qabs_tls}).
    \item 
    The physical properties of the amorphous dust expected from the results of SED fitting for the Perseus MC using the SP model differ by several orders of magnitude from typical values obtained from laboratory measurements of amorphous materials (see Tab.\,\ref{tab:params_est}).
\end{enumerate}

The main purpose of this paper is to derive the amorphous dust emission SED based on the SP model. Detailed comparisons with astronomical observations and laboratory measurements will be conducted in future work. We did not impose any restrictions on the fitting parameters in the SED fitting of the Perseus MC. Comparing the SP model with observational data and imposing the results of laboratory measurements as necessary conditions to verify the rationality of the SP model is essential. For this reason, it is necessary to investigate the reproducibility of the laboratory measurements of the optical constants for amorphous materials (e.g., \citealp{Demyk+22}).
Physical quantities of amorphous dust estimated from the SED fitting for the Perseus MC were inconsistent with the results of the measurement of the thermal properties of amorphous materials. One possible interpretation of these results is that the amorphous materials composing interstellar dust could differ in chemical composition or physical structure from the laboratory. The development of theoretical research that utilizes molecular dynamics simulations is paramount for estimating the chemical composition and physical structure of amorphous materials in interstellar dust with unknown properties. In this paper, we have focused on the long wavelength side of the amorphous dust emission, so we have only dealt with the TLS of the minute energy differences originating from the DWP. On the other hand, the effects of state transitions to higher-level states and the anharmonicity of the single-well potential will come into play in the short wavelength. By taking these effects into account, it is expected that a unified understanding of interstellar dust will be achieved by modeling the optical properties of amorphous dust over a wide range of wavelengths, from ultraviolet to microwave, using multi-wavelength observational data.
\begin{acknowledgements}
    We would like to thank the referees for their constructive comments and suggestions, which have greatly improved this paper. 
    We are also deeply grateful to Itsuki Sakon for many valuable discussions throughout the development of the paper.
    MN acknowledges support from JSPS KAKENHI Grant Number JP22J00388.
    This work was supported by MEXT KAKENHI Grant Number JP20KK0065 and JSPS KAKENHI Grant Number JP24K17085.
\end{acknowledgements}

\bibliography{reference}{}
\bibliographystyle{aasjournal}

\end{document}